# [Invited] Advances in All-silicon Waveguides for Terahertz Integration

Nguyen H. NGO, Weijie GAO, and Masayuki FUJITA

Graduate School of Engineering Science, The University of Osaka
1-3 Machikaneyama, Toyonaka-shi, Osaka, 560-8531 Japan
E-mail: ngo.h-nguyen.es@osaka-u.ac.jp

**Abstract** In chip-to-chip communication, terahertz waves provide a promising approach to achieve high data capacity with improved energy efficiency, effectively bridging the gap between electrical and optical domains. By realizing this potential, all-dielectric waveguides have emerged as high-speed interconnects, offering broad bandwidth and low-loss characteristics while being compatible with planar integration. This paper reviews different types of terahertz all-dielectric waveguides, including photonic-crystal and effective-medium-clad structures, and highlights recent progress in implementing these platforms to develop various passive functional devices, paving the way for high-speed operation and large-scale integration in terahertz systems.

## 1. Introduction

Terahertz (THz) waves, extending from 0.1 to 10 THz, have attracted growing interest for next-generation communication networks, as well as for imaging and sensing applications. THz communications offer enormously available bandwidth, enabling ultra-high data rates beyond those of current millimeter-wave systems. Recent demonstrations have achieved over 100 Gbit/s with a single channel at the 300-GHz band and 1 Tbit/s with multiple carrier frequencies in the WR-1.5 band (500–750 GHz) [1], underscoring the potential of THz links for future 6G networks. Besides, THz waves interact and penetrate dielectric materials, which opens unique opportunities in spectroscopy and nondestructive sensing and imaging [2]. To fully exploit the THz band, compact and efficient THz sources are essential. Optical-based generation usually requires complex system integration and low DC-to-RF efficiency. Conversely, electronics-based oscillators have higher energy efficiency and larger power density, yet fundamental oscillation frequencies are limited to ~2 THz with significantly reduced output powers [3]. Therefore, preserving power is crucial for high-speed inter-/intra-chip communication systems, enabling the development of low-loss THz interconnects.

Another challenge in conventional THz systems is the high propagation loss of metallic-based waveguides at these frequencies. Traditional microwave-inspired guiding structures (e.g., metallic rectangular waveguides or coaxial cable) suffer severe ohmic losses in the THz regime. Although metallic hollow waveguides can achieve low attenuation (theoretically ~2 dB/cm up to 1 THz [4]) with most waves propagating in the air core, they are not applicable to build integrated circuits for their non-planar geometry. THz coplanar waveguides are suitable for planar integration; however, metal-dependent structures also exhibit larger attenuation (e.g., ~35 dB/cm at 750 GHz [5]) compared to metallic hollow waveguides. To eliminate the ohmic loss, dielectric waveguides (e.g., silicon-on-insulator or all-silicon waveguides) have emerged for efficient THz signal transmission, where high-resistivity silicon waveguides have demonstrated ~2 order-of-magnitude lower attenuation compared to their metallic counterparts at 1 THz [6].

Planar waveguides are therefore considered as a candidate for THz integrated systems. A simple high-index ribbon waveguide surrounded by air confines waves within the core through total internal reflection (TIR). By introducing effective-medium or photonic crystal cladding, a substrate-less platform is realized for multiple purposes, including the ease of waveguide handling. Notably, this substrate-less platform is compatible with microfabrication processes, such as deep reactive-ion etching, thereby lowering fabrication complexity and eliminating the excess loss inherent in platforms like silicon-on-glass. In addition to low loss, all-dielectric waveguides can offer broad operational bandwidth with low dispersion, which is critical for high-speed data transmission. Besides, all-dielectric waveguides provide a foundation for various passive components (e.g., filters, couplers, antennas), highlighting their potential for practical THz front-end devices. In this paper, we present the fundamentals of all-dielectric planar waveguides, being categorized based on their guiding mechanism, and review our recent advances in THz integrated components developed from these waveguides.

## 2. Fundamentals of all-dielectric planar waveguides

Dielectric waveguides confine THz waves through TIR or photonic bandgap (PBG) effects. They serve as the fundamental building blocks for passive components such as splitters, resonators, and mode converters. Thanks to these advantages, all-dielectric waveguides are poised to become key building blocks for THz integrated circuits and systems, providing the low-loss interconnects and functional elements needed for practical THz communications, sensing, and imaging devices.

Depending on the guiding mechanism and cladding geometry, THz dielectric waveguides can be categorized into (a) air clad, (b) photonic-crystal clad, (c) effective-medium clad, and (d) slot-based waveguides, as illustrated in Fig. 1.

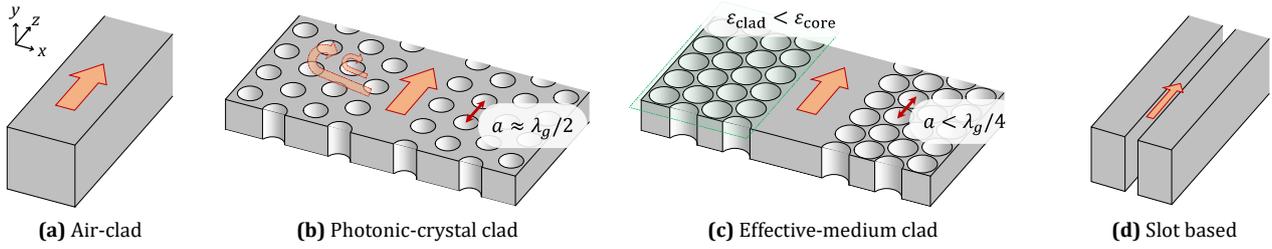

*Figure 1 Various types of all-dielectric waveguide based on wave confinement mechanisms: (a) all-air clad, (b) photonic-crystal clad, (c) effective-medium clad, (d) slot based. Waves propagate along the z-axis. a: lattice constant, $\lambda_g$: guided wavelength at central frequency.*

## 2.1. Air clad

A simple yet efficient waveguide is a ribbon waveguide. Owing to the etching process, the cross-section of the waveguide has a rectangular shape. Since the silicon core has high refractive index ($n \approx 3.4$) and is cladded by air with low index ($n = 1$), the wave is strongly confined by TIR in both in-plane (x-z) and out-of-plane (y-z) directions. A number of silicon ribbon waveguides developed for frequency bands from 100 GHz up to 1 THz with high transmittances have been reported [4]. Since the bare guiding channel has a small cross-section (<1×1 mm$^2$ for sub-THz bands) without supporting structures, it is impractical to manipulate the waveguide for accommodating with other components. Therefore, tethered frame or low-index claddings are introduced. However, these claddings will also cause the wave interference to the structure, and there will be a trade-off between the transmittance and physical strength of the waveguide [7], [8]. In addition to silicon, 3D-printed alumina is a promising technology, as the relative permittivity of materials can be arbitrarily controlled [9].

It is worth noting that a pair of metallic hollow-waveguide ports are attached to both ends of each silicon waveguide to excite the fundamental TE (transverse-electric) or TM (transverse-magnetic) mode. In order to mitigate impedance mismatch between the metal waveguide and the silicon core, tapered spikes are incorporated at both terminations of the dielectric waveguides. These tapers gradually modify the effective refractive index and modal field distribution, thereby reducing reflection at the interface and allowing high transmittance.

## 2.2. Photonic-crystal clad

Photonic crystal (PC) is an artificial periodic structure. The structure exhibits a PBG effect, where the light with frequency within the bandgap cannot pass through the structure owing to the destructive interferences. A lattice of air hole is formed where the period is approximately half of guided wavelength [10]. Introducing a line defect creating a guided mode within this bandgap, enabling the transmission along the defect. Consequently, PC claddings offer both wave confinement and mechanical support. Recently, valley photonic crystal (VPC) waveguides achieve topological protection by breaking inversion symmetry of the unit cell, which opens a valley-dependent bandgap [11]. By interfacing two PCs with opposite valley Chern numbers, a topological edge channel is formed along the interface. This guided mode exhibits strong suppression of backscattering, robustness to fabrication deviations, and high transmission through sharp bends.

## 2.3. Effective-medium clad

Since the PC and VPC waveguides rely on the PBG effect, the bandwidth is limited by around ~10%. In some applications that require higher bandwidth (e.g., communications) the choice of PC-based is not sufficient. By employing hexagonal lattice similar to that of PC, the period of air holes is determined to be much smaller than the guided wavelength. Thus, the effective media can be treated as a homogeneous material with lower refractive index than that of the guided core, and the wave is well confined within the core based on TIR [12]. Additionally, the advantage of effective-medium structure is the control of the refractive index. By engineering the fill factor of air, the cladding achieves a tailorable anisotropic permittivity tensor based on Maxwell–Garnett approximation. The operation bandwidth of the effective-medium waveguide is ~40%. To further reduce the evanescent leakage to the effective media, extreme skin-depth claddings were introduced to manipulate the permittivity tensor for maximizing the in-plane anisotropy ratio [13], which is potentially applied for densely integrated devices.

## 2.4. Slot-based waveguide

In previous waveguides, the waves are confined within a high-index material (silicon). On the other hand, the slot waveguide confines the wave within a minute air gap sandwiched by a high-index material. The mechanism is based on the Maxwell equation, stating that the normal component of the electric flux density must be continuous at the boundary, i.e., $\varepsilon_{Si}E_{n,Si} = \varepsilon_{air}E_{n,air}$. Therefore, the E-field right at the silicon-air boundary in the air gap is ~12 ($\varepsilon_{Si}/\varepsilon_{air}$) times that in silicon. Such slot-based principles were applied for both out-of-plane and in-plane configurations [14], [15]. When feeding the waveguide from a metallic hollow waveguide, a curvature is introduced at the termination of the silicon beam is employed to reduce the index and impedance mismatches at the interface while eliminating the conventional tapered spikes, enabling efficient taperless coupling [16].

## 3. Advances in Terahertz All-silicon Integrated Components

Building upon the fundamental dielectric waveguides introduced earlier, various THz passive devices have been developed to enable advanced communication, sensing, and imaging functionalities. These devices leverage low loss and compact features, allowing scalable integration on all-dielectric substrates. This section summarizes representative state-of-the-art components and applications, including their designs and operation principles, to future THz systems.

## 3.1. Valley photonic crystal triplexer

The triplexer was developed to (de)multiplex signal on a VPC platforms to reduce the backscattering with a relatively high isolation [17]. Starting with a basic bearded interface, the propagation mode exists in the whole bandgap (320–375 GHz). Then, a diplexer is developed by employing a frequency filter formed by a zigzag-interface with a certain distance. By optimizing the size of zigzag-interface triangles, the propagation mode exists and the filter

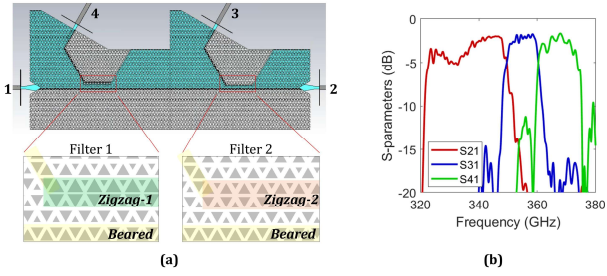

*Figure 2 (a) Low-pass filters formed by zigzag-interface with different size in VPC triplexer. (b) Simulated S-parameters.*

functions as a low-pass filter with different cutoff frequency. High-frequency waves cannot travel and be reflected to a sharp-bend bearded interface. As a result, a triplexer is done when combining two diplexers while fully utilizing the whole frequency band determined by the VPC's bandgap, as shown in Fig. 2.

### 3.2. Photonic-crystal power combiner

A PC Y-branch power combiner [18] has been implemented to coherently increase the power of multiple THz sources, such as resonant tunneling diodes (RTDs) [3]. The device uses a bifurcation structure to merge the incoming signals from two RTDs. Owing to the mutual injection locking effect, the RTDs can be synchronized when their oscillation frequencies are close to each other. By optimizing the air holes at the Y-junction, the transmittance reaches ~0.71 ($1/\sqrt{2}$), which is theoretically maximum. In the Y-branch, the mutual coupling between two RTDs equals to 0.5, ensuring that the RTD can interact with the other one. Experimental results confirm that both RTDs are locked and power enhancement is 4 times compared to a standalone RTD in the structure, shown in Fig. 3. This photonic-crystal structure is efficient for on-chip transmitters where higher power levels are required with a limited power of single THz sources.

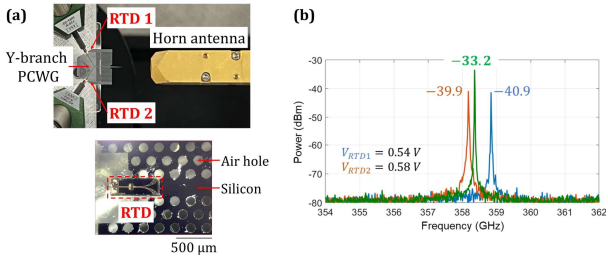

*Figure 3 (a) Two RTDs couple to PC Y-branch waveguide. Output power is collected by using horn antenna. (b) Spectra of standalone RTDs (red and blue). When simultaneously biasing both RTDs, combined power (green) is enhanced.*

### 3.3. Polarization multiplexer

An effective-medium based polarization multiplexer was developed to simultaneously guide two orthogonal polarization modes, thereby doubling the data throughput within a single physical path [19]. The main feature of this device is its broad bandwidth, high polarization extinction ration, and low loss. Specifically, the developed multiplexer supports TE and TM modes, both are excited at port 1, as shown in Fig. 4. The polarization paths are divided into two parts. The $E_{11}^y$-path (for out-of-plane polarization, to port 3) consists of two tapered directional couplers connected by a 90°-curved bend. A small air gap between two couplers with sufficient long coupling length ensures high power transfer while rejecting the $E_{11}^x$ mode. In contrast, the $E_{11}^x$-path (for in-plane polarization, to port 2) comprises two tapered waveguides bridged by a thin straight waveguide with a wide gap. A combination of this wide coupling gap filled by high-index effective-medium material enables leakages of the $E_{11}^x$ mode from the taper to the narrow bridge. Importantly, the utilization of effective-medium cladding further increases the polarization extinction ratios of more than 20 dB for both modes. The insertion losses for two paths are below 1 dB in the whole WR-3.4 (220–330 GHz) band. The device successfully demonstrated an aggregated data rate of 155 Gbit/s, showing capability for broadband THz communications for doubling data throughput while maintaining a compact on-chip footprint.

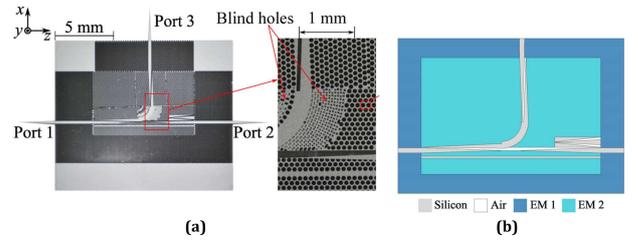

*Figure 4 (a) Fabricated sample and (b) segmented regions with different permittivity tensors of polarization multiplexer. Copyright 2024 The Authors, licensed under a Creative Commons Attribution-NonCommercial License. (https://creativecommons.org/licenses/by-nc/4.0/)*

### 3.4. Spikeless-interfaced waveguide

To overcome the mechanical fragility associated with long tapered silicon spikes, a spikeless interface has been introduced for coupling a dielectric waveguide to a standard hollow metallic waveguide [20]. This design employs a slot waveguide as a coupler without any protruding section to the hollow core. To minimize the impedance mismatch between air and silicon at the interface, a tapered transition region is optimized to maintain high transmittance in the whole WR-2.2 band (330–500 GHz). To connect two beams of the slot waveguide, an array of sub-wavelength elliptical air holes is used to form a low-index anisotropic material, as shown in Fig. 5. In practice, this waveguide significantly simplifies alignment and improves durability during assembly, making it well suited for practical THz module packaging and deployment. By extending the operation frequency to 600 GHz, the waveguide can support communications for 14 channels with an aggregated data rate of more than 800 Gbit/s, which is compatible for ultra-high-speed integrated circuit.

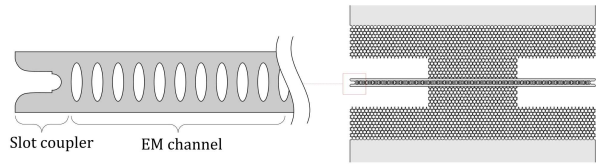

*Figure 5 Spikeless waveguide based on slot coupler and effective-medium channel. Period of elliptical air holes is 50 μm.*

### 3.5. Truncated rod for imaging

For sensing and imaging applications, near-field probes based on dielectric waveguides have been developed to achieve enhanced spatial resolution and material contrast [21], as illustrated in Fig. 6. In particular, truncated silicon waveguides provide strong field localization at the tip, enabling high-resolution imaging with compact hardware. These probes have been applied to semiconductor inspection, conductivity mapping, and broadband material classification, illustrating their utility as cost-effective sensing elements compatible with silicon-based THz platforms. Their simple fabrication and robust performance make them attractive for integrated non-destructive evaluation systems.

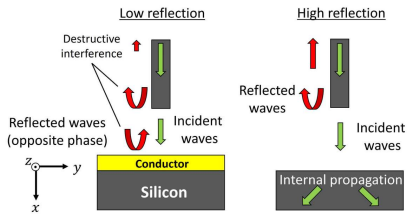

*Figure 6 Concept of truncated silicon waveguide.
Copyright 2025 The Authors, licensed under a Creative Commons Attribution-NonCommercial-NoDerivatives 4.0 License (https://creativecommons.org/licenses/by-nc-nd/4.0/).*

## 4. Conclusion

THz all-silicon waveguides have emerged as an efficient platform owing to their extremely low propagation loss (~0.05 dB/cm). The guiding mechanisms are categorized based on TIR, PBG, and slot confinement effects, providing strong field confinement in the microfabricated structure. By leveraging these mechanisms, fundamental waveguide platforms can be extended into functional passive components such as polarization/frequency multiplexers and power combiners/splitters, enabling ultra-high-speed data transmission and compact solutions for non-destructive sensing and imaging across the THz regime.

Despite these extensive demonstrations, the transition of dielectric-waveguide-based technology into commercially deployable systems remains limited. Heterogeneous integration of developed components into multi-functional systems compatible with existing electronic/photonic ecosystems (e.g., fiber optics, antenna) will be essential for future 6G-and-beyond applications. Our next direction is the hybrid integration of active THz sources and detectors, such as III–V diodes, directly onto silicon waveguide platforms to realize compact and portable modules. Ultimately, the convergence of passive silicon waveguides and active devices is expected to transform dielectric-waveguide technology from laboratory-scale prototypes into practical platforms for real-world applications.

## Acknowledgements

This work was supported in part by CREST, JST (JPMJCR21C4) and KAKENHI (24H00031).